# CT RECONSTRUCTION WITH PDF: PARAMETER-DEPENDENT FRAMEWORK FOR MULTIPLE SCANNING GEOMETRIES AND DOSE LEVELS

*Wenjun Xia[1], Zexin Lu[1], Yongqiang Huang[1], Yan Liu[2], Hu Chen[1], Jiliu Zhou[1], Senior Member, IEEE, Yi Zhang[1,*], Senior Member, IEEE*

[1] College of Computer Science, Sichuan University, Chengdu 610065, China
[2] College of Electrical Engineering, Sichuan University, Chengdu 610065, China

**ABSTRACT**

Current mainstream of CT reconstruction methods based on deep learning usually needs to fix the scanning geometry and dose level, which will significantly aggravate the training cost and need more training data for clinical application. In this paper, we propose a parameter-dependent framework (PDF) which trains data with multiple scanning geometries and dose levels simultaneously. In the proposed PDF, the geometry and dose level are parameterized and fed into two multi-layer perceptrons (MLPs). The MLPs are leveraged to modulate the feature maps of CT reconstruction network, which condition the network outputs on different scanning geometries and dose levels. The experiments show that our proposed method can obtain competing performance similar to the original network trained with specific geometry and dose level, which can efficiently save the extra training cost for multiple scanning geometries and dose levels.

## 1. INTRODUCTION

Computed tomography (CT) is widely used for clinical diagnosis. However, the patients suffer from radiation risk during scanning. To protect the health of patients, the exposure level can be reduced by switching the voltage or current of the x-ray tube or reducing the sampling numbers, which are referred to low-dose and sparse-view technique, respectively. However, both methods will degrade the imaging quality. Therefore, how to improve the reconstruction quality with a lower radiation dose has always been a major topic.

With the rapid development of deep learning, learning-based methods have become the mainstream of CT reconstruction. These methods can be roughly classified into two categories: 1) the image-to-image method and 2) the data-to-image method. The first kind of methods can remove the noise and artifacts by directly learning a mapping from the noisy image to the clean image, such as [1-4]. However, since these methods excludes the measured data in the computation, the data consistency cannot be well guaranteed. The second kind of methods can address this challenge. By feeding the projection data into the network simultaneously, the reconstruction can be progressively corrected and has a higher accuracy than post-processing methods. These methods implement the mapping from data domain to image domain. There are two representative ways to achieve this goal. [5, 6] learned a full connected network to perform the mapping. In [7, 8], the system matrix is embedded into the network to calculate the projection and back-projection. However, all these methods mentioned above need to fix the geometry and dose level. Since different scanners probably have different geometries, and even the same scanner may use different geometries and dose levels for different scanning tasks, we have to train multiple networks for different geometries or dose levels, which will significantly aggravate the training cost and need more training data.

Inspired by [9], we propose a parameter-dependent framework (PDF) to process the data with different scanning geometries and dose levels in a unified network. This framework can be applied for most current learning-based CT reconstruction methods, including both image-to-image and data-to-image methods. Arbitrary reconstruction network can be chosen as the backbone. First, the geometry and dose level are parameterized and fed into two multi-layer perceptrons (MLPs). Then, the output of MLPs are used to condition the reconstruction network by modulating the feature maps of the network. For the data-to-image model, the data accompanied by the geometry parameters are fed into the module to calculate the projection and back-projection in real time, achieving the dynamic mapping from data domain to image domain.

The rest of paper is organized as follows. In the next section, we present our proposed method in detail. In the third section, the results of the comparative experiments are demonstrated. The last section provides the conclusion.

## 2. METHODOLOGY

Considering the existing deep learning models
$$x^* = F(x, y, \theta) \tag{1}$$

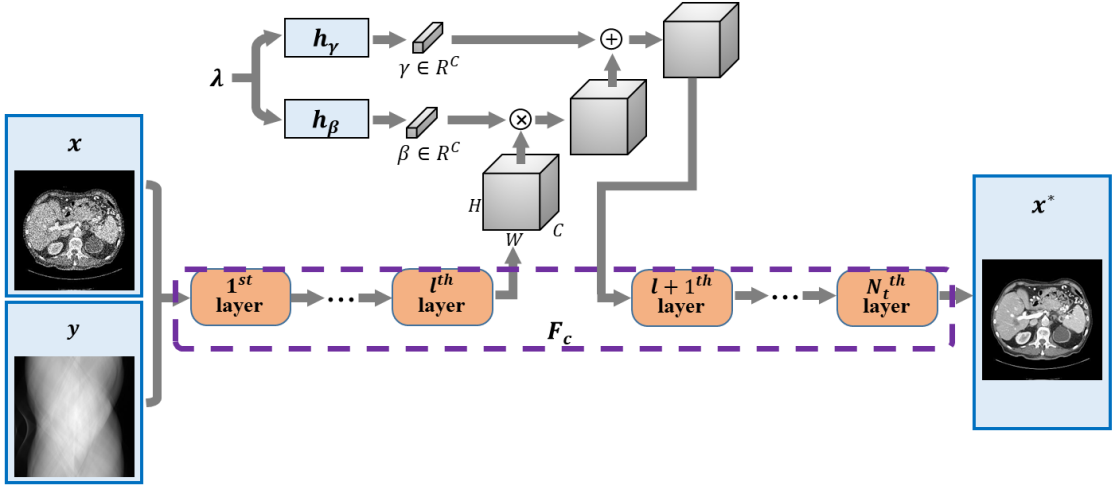

**Fig. 1.** Modulation of $l$-th layer of the reconstruction network.

where $x \in R^N$ is the noisy image, which is reconstructed with the measured projection data $y \in R^M$ with a specific geometry $\Sigma$ and X-ray's intensity $I_0$, which is associated with the dose level. $F$ is the reconstruction network with trained parameters $\theta$, and $x^*$ is the prediction of network. In this model, the trained parameters can only process the data with the specific geometry $\Sigma$ and dose level $I_0$:

$$\theta^*_{\Sigma, I_0} = \arg\min_\theta \mathrm{E}\{L(F(x,y,\theta), \hat{x})\} \quad (2)$$

where $\hat{x}$ denotes the label and $L()$ is the loss function. For the image-to-image methods, the network is trained with paired label image and noisy image with specific noise level determined by both $\Sigma$ and $I_0$. Once the noise level changes, the results will be unpredictable. For the data-to-image methods, since the network is usually designed for specific $\Sigma$, such as AUTOMAP (fixed size of the input measured data) and LEARN (fixed system matrix), the network cannot even work. To overcome the aforementioned challenge, we parameterize the geometry and dose level, and feed it into the network, resulting the following framework:

$$x^* = F_c(x, y, \theta, \lambda) \quad (3)$$

where $F_c$ is the conditional network, and $\lambda = P(\Sigma, I_0)$ is the feature vector obtained by parameterizing the geometry and dose level. The general parameters for different scanning geometries and dose levels can be optimized as:

$$\theta^* = \arg\min_\theta \mathrm{E}\{L(F_c(x,y,\theta,\lambda), \hat{x})\} \quad (4)$$

In Eqs. (3) and (4), $F_c$ is a conditional network, whose performance depends on the feature vector $\lambda$. In this work, the feature-wise linear modulation (FiLM) [10] is leveraged to condition the network. The feature vector $\lambda$ is utilized to modulate all the convolutional layers of $F$. Assume that the feature map extracted from the $l$-th layer is $f^l \in R^{W \times H \times C}$, which has a spatial resolution of $W \times H$ and $C$ channels. $\lambda$ is fed into two MLPs $h_\beta$ and $h_\gamma$, then two posterior parameters $\beta \in R^C$ and $\gamma \in R^C$ were obtained. $f^l$ is multiplied channel-wise by $\beta$ and added to $\gamma$ to get the modulated feature map $\tilde{f}^l$:

$$\tilde{f}^l_{ijk} = \beta_k f^l_{ijk} + \gamma_k, \quad \beta = h_\beta(\lambda), \gamma = h_\gamma(\lambda) \quad (5)$$

TABLE I Different combinations of scanning geometries and dose levels

| | #1 | #2 | #3 | #4 | #5 |
|---|---|---|---|---|---|
| Number of views | 1024 | 88 | 1024 | 128 | 108 |
| Number of detector bins | 512 | 768 | 768 | 512 | 512 |
| Length of pixel (mm) | 0.66 | 0.78 | 1.0 | 1.2 | 0.5 |
| Length of detector bin (mm) | 0.72 | 0.58 | 0.62 | 1.4 | 0.4 |
| Distance between source and rotation center (mm) | 250 | 350 | 500 | 500 | 400 |
| Distance between detector and rotation center (mm) | 250 | 300 | 400 | 500 | 200 |
| Intensity of X-rays | 1e5 | 1e6 | 5e4 | 2.5e5 | 5e5 |

Fig. 1 illustrates the modulation of the $l$-th layer of the reconstruction network. Particularly, if there are feature maps with different channel numbers in the reconstruction network, multiple sets of MLPs are needed. In fact, the modulation is essentially a transformation between different distributions. Assume that the learned feature maps follow a normal distribution $p(f) \sim N(\mu, \sigma^2)$. We use two MLPs to learn the posterior probability distribution of feature maps $p(f|\lambda) \sim N(\beta\mu + \gamma, (\beta\sigma)^2)$. Then Eq. (5) becomes the process of distribution transformation.

## 3. EXPERIMENTS AND RESULTS

To evaluate the performance of proposed PDF, the dataset "*the 2016 NIH-AAPM-Mayo Clinic Low-Dose CT Grand Challenge*", which has 5936 full-dose CT images from 10 patients, was used in our experiments. We randomly selected 400 images from eight patients as the training set. These images are randomly divided into five groups on average, and each group used different scanning geometries and dose levels to simulate the projection data. The parameters of the scanning geometries and dose levels are shown in TABLE I. $I_0 = 1e^6$ can be treated as the normal dose. The cases that the number of sampling views less than 1024 can be regarded as sparse-view. Meanwhile, we randomly selected 100 images from the remaining two patients as the testing set. While testing, the testing images were simulated with the five scanning geometries and dose levels and then reconstructed with the trained network. To simulate a realistic clinical environment, Poisson noise and electronic noise were added into the measured projection data as [11]

$$y = \ln \frac{I_0}{\text{Poisson}(I_0 \exp(-\hat{y})) + \text{Normal}(0, \sigma_e^2)} \quad (6)$$

where $\hat{y}$ represents the noisy-free projection, and $\sigma_e^2$ is the variance of electronic noise. In our experiments, we fixed electronic noise variance at $\sigma_e^2 = 10$.

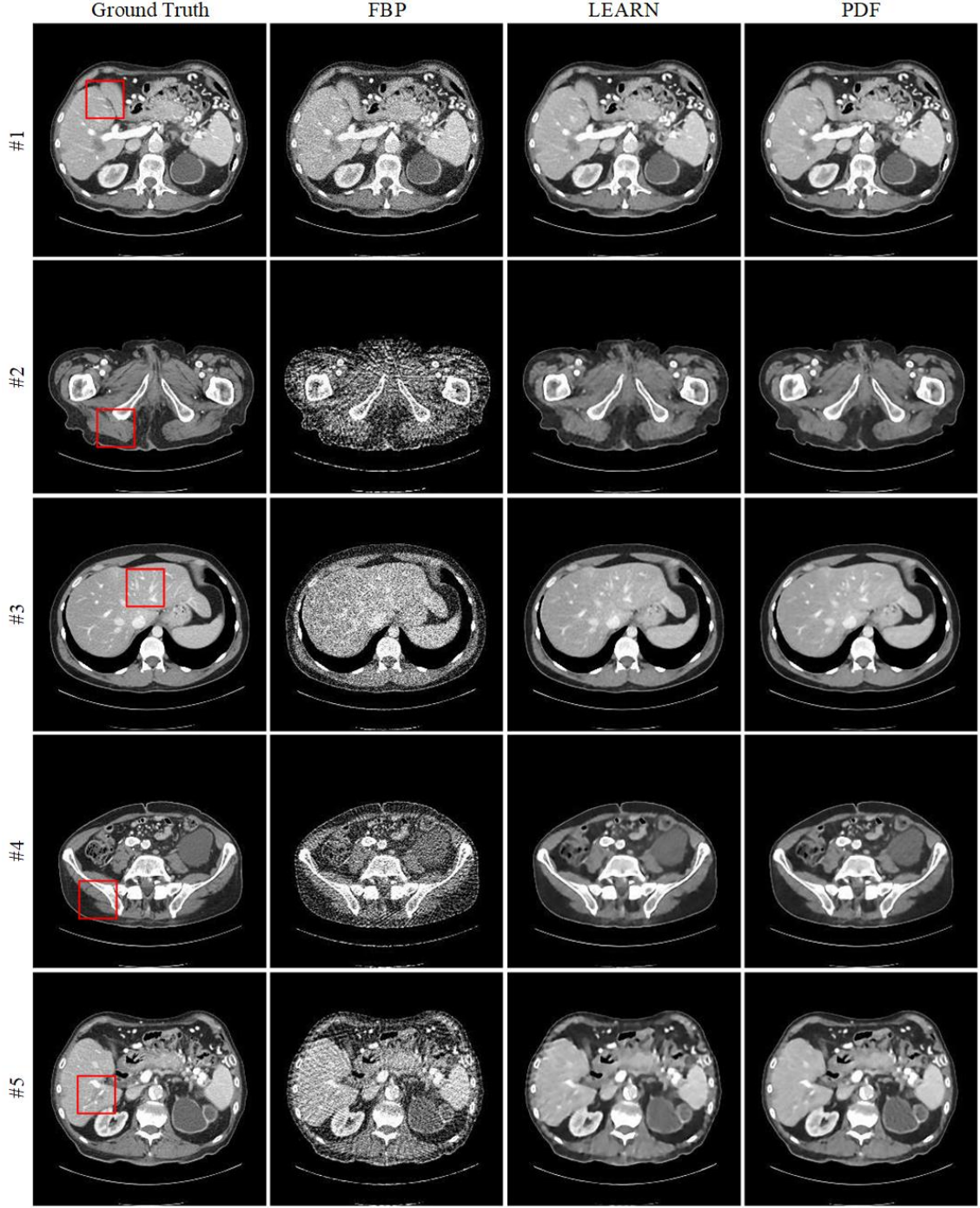

**Fig. 2.** The results of different scanning geometries and dose levels by LEARN and PDF. The display window is [-160, 240] HU.

In our experiments, LEARN [7] was selected as the reconstruction network. We trained LEARN-based PDF and LEARN separately. PDF was trained with the mixed data. LEARN was trained with the five groups of data separately and five models corresponding to five scanning geometries and dose levels were obtained. In LEARN, the system matrix is embedded in the network. To accelerate LEARN-based PDF, CUDA was used to code the operations of projection and back-projection. The matched data and geometry are fed into the module to get the projection or back-projection. The mean square error (MSE) is adopted as the loss function. And the training epochs of both methods were 200.

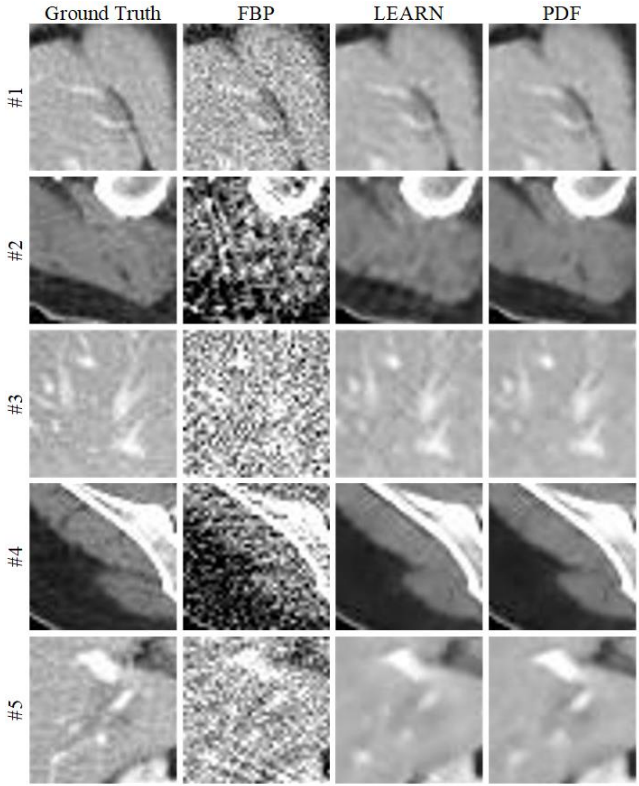

**Fig. 3.** Zoomed regions indicated in Fig. 2.

Fig. 2 shows the results of different scanning geometries and dose levels by LEARN and PDF. #1 and #3 are the low-dose cases, and #2, #4 and #5 are the sparse-view cases. In the results of low-dose cases, PDF can achieve performance close to LEARN. Both methods can remove the noise and recover the structures effectively. Specially, the metastases are well maintained. However, in the sparse-view cases, LEARN cannot eliminate the artifacts thoroughly. On the contrary, the artifacts are well suppressed in the results of PDF. Although the amount of measured data of a specific geometry and dose level in PDF is the same as in LEARN, training with mixed data makes PDF more robust, which improve the performance for sparse-view data. To better observe the details using different methods, the regions of interest (ROIs) indicated by red boxes in Fig. 2 are magnified in Fig. 3. It can be seen that PDF achieved competing performance to LEARN. In the results of PDF, the structures are well preserved, especially near the clinically important lesion shown in #5 of Fig. 3. The statistical results of the whole testing set are shown in TABLE II, which gives the means and standard deviations (SDs) of both PSNR and SSIM. It can be seen that for low-dose cases, our proposed PDF obtained similar performance to LEARN. This performance is acceptable when training efficiency is greatly improved. What is interesting that the performance of PDF is much better than LEARN for sparse-view data. This proves that our framework can increase the data diversity, which can significantly improve the performance when data is insufficient.

TABLE II Quantitative results (Mean±SD) of different scanning geometries and dose levels

|    | LEARN | | PDF | |
|----|-------|------|------|------|
|    | PSNR  | SSIM | PSNR | SSIM |
| #1 | 35.22±0.63 | 0.9538±0.0085 | 34.64±0.64 | 0.9493±0.0100 |
| #2 | 26.69±1.07 | 0.8632±0.0208 | 30.01±1.12 | 0.9042±0.0188 |
| #3 | 29.02±1.04 | 0.9274±0.0120 | 28.90±1.61 | 0.9307±0.0119 |
| #4 | 27.89±0.69 | 0.8755±0.0161 | 29.69±0.87 | 0.9039±0.0168 |
| #5 | 27.00±1.00 | 0.8404±0.0224 | 30.79±0.98 | 0.9137±0.0157 |

## 4. CONCLUSIONS

In this paper, we propose a novel model to perform CT reconstruction with data with different scanning geometries and dose levels. It can be concluded that the performance of the proposed general framework can achieve competing performance to the model trained with specific geometry and dose level. This performance is acceptable when we save extra training cost for multiple reconstruction networks. Especially, our model can improve the diversity of data efficiently, which significantly improves the performance of reconstruction network while the training data is insufficient.